\RequirePackage{lineno}
\documentclass[%
 aip,
 amsmath,amssymb,
preprint,%
]{revtex4-1}

\usepackage{graphicx}
\usepackage{float}
\usepackage{dcolumn}
\usepackage{bm}
\usepackage{hyperref}
\hypersetup{
    colorlinks=true,
    linkcolor=blue,
    filecolor=magenta,      
    urlcolor=blue,
    citecolor=blue,
}
\usepackage{color}

\begin{document}


\title{Electronic and Optical Properties of Ultrawide Bandgap Perovskite Semiconductors via First Principles Calculations}%

\author{Radi A. Jishi}
\affiliation{Department of Physics and Astronomy, California State University, Los Angeles, California 90032, USA}
\author{Robert J. Appleton}
\affiliation{School of Materials Engineering and Birck Nanotechnology Center, Purdue University, West Lafayette, Indiana 47907, USA}
\author{David M. Guzman}
 \email[Corresponding Author: ]{davgumo@me.com}
\affiliation{School of Materials Engineering and Birck Nanotechnology Center, Purdue University, West Lafayette, Indiana 47907, USA}

\date{\today}

\begin{abstract}
Recent research in ultrawide-bandgap (UWBG) semiconductors has focused on traditional materials such as Ga$_2$O$_3$, AlGaN, AlN, cubic BN, and diamond; however some materials exhibiting single perovskite structure have been known to yield bandgaps above 3.4 eV, such as BaZrO$_3$. In this work we propose two novel materials to be added to the family of UWBG semiconductors: Ba$_2$CaTeO$_6$ exhibiting a double perovskite structure and Ba$_2$K$_2$Te$_2$O$_9$ with a triple perovskite structure. Using first principles hybrid functional calculations we predict the bandgaps of all the studied systems to be above 4.5 eV with strong optical absorption in the ultraviolet region. Furthermore, we show that holes have a tendency to get trapped through lattice distortions in the vicinity of oxygen atoms with average trapping energy of 0.25 eV, potentially preventing the enhancement of \textit{p}-type conductivity through traditional chemical doping.   
%
\end{abstract}


\maketitle

Technological developments in electronics, photonics, electro- and photo-chemistry have been based in traditional materials such Ge, Si, and III-Vs compounds \cite{RevModPhys.73.783}, all generally characterized by bandgaps under 2.3 eV. However, in the last 10 years the wide bandgap semiconductor InGaN became the second most important material, behind Si\cite{doi:10.1002/andp.201570058}, because of its application in solid state lighting\cite{nakamura}, which has rapidly changed how the world makes use of light sources\cite{doi:10.1002/pssa.201026349}. 
    
Ultrawide bandgap (UWBG) semiconductors are materials with a bandgap wider than 3.4 eV, the bandgap of GaN. These materials are at the forefront of semiconductor research and will be the building blocks for new devices and applications\cite{doi:10.1002/aelm.201600501}. Some figures of merit for device performance, such as the Baliga figure of merit\cite{doi:10.1063/1.331646} for low-frequency power switches, and the Johnson figure of merit\cite{rca} for high-frequency applications, have a nonlinear dependence on the bandgap. Replacing conventional semiconductors with these UWBG semiconductors in areas such as high-power and RF-electronics, as well as deep-UV optoelectronics, could result in rapid improvements in device performance.
    
Acceptor doping has been difficult to achieve in UWBG semiconductors mainly due to the tendency of holes to self-trap\cite{PhysRevB.85.081109,doi:10.1063/1.5036750} and the presence of compensating defects. However, it was recently demonstrated that by pinning the Fermi level in AlGaN/AlN away from the valence band edge during epitaxy, it was possible to realize a reduction in the formation energy of substitutional Mg-dopant along with an increase in the formation energy of compensating defects, leading to enhanced p-type doping\cite{PhysRevMaterials.3.053401}.
    
Though most of the research on UWBG semiconductors has focused on AlGaN/AlN, diamond, and Ga$_2$O$_3$, the domain of UWBG semiconductors is not restricted to those materials. Other materials have also been explored, such as the ternary oxides MgGa$_2$O$_4$\cite{doi:10.1002/pssa.201431835}  and  ZnGa$_2$O$_4$\cite{doi:10.1021/acs.cgd.9b01669},  the ternary nitrides  MgSiN$_2$  and ZnSiN$_2$\cite{Endo1992}, the alloys $\alpha$-(AlGa)$_2$O$_3$\cite{Jena}, and two-dimensional GaN realized via graphene encapsulation\cite{10.1038/nmat4742}.

As technology evolves, pushing the intrinsic performance boundaries of materials to their limits, we face the need to find novel alternative materials to enable the continuation of technological development. Research in the relatively new field of UWBG semiconductors continues to focus on traditional aluminum- and gallium-based materials; however, prototypical structures taken mostly from energy harvesting applications, such as barium-based perovskites, show great potential to be included in the UWBG semiconductors category, bringing a new class of chemistries and structures into the field.

\begin{figure*}[ht]
\centering
\includegraphics[scale=0.5]{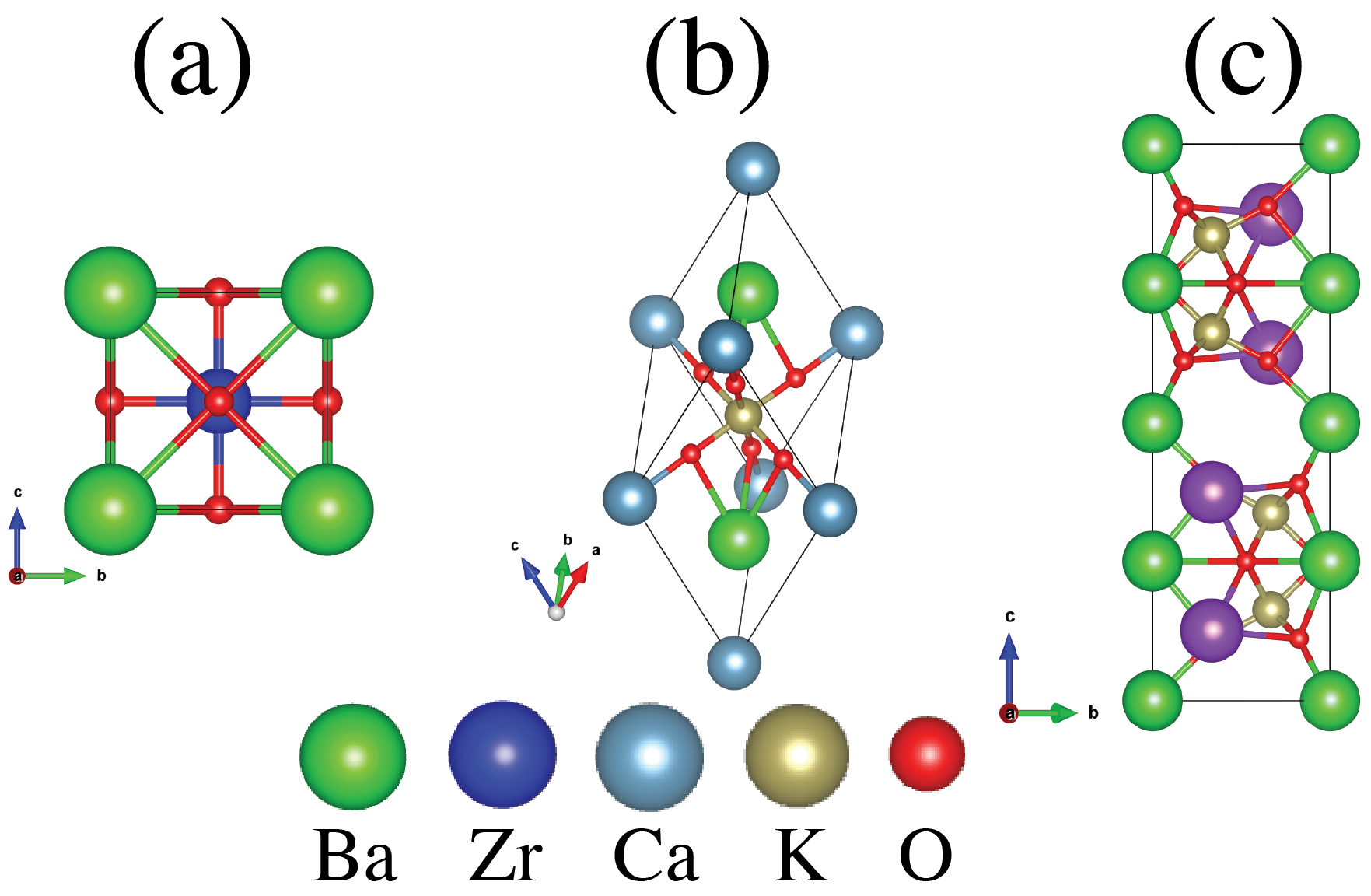}
\caption{Primitive cells for (a) simple cubic BaZrO$_3$, (b) face-centered cubic Ba$_2$CaTeO$_6$, and (c) hexagonal Ba$_2$K$_2$Te$_2$O$_9$}
\label{fig:struct}
\end{figure*}

In this letter, we use first-principles calculations to study the electronic and optical properties of BaZrO$_3$, Ba$_2$CaTeO$_6$, and Ba$_2$K$_2$Te$_2$O$_9$, three crystals of the perovskite type. Using hybrid functionals, we show that these are UWBG semiconductors with band gaps exceeding 4.9 eV and with significant absorption in the deep-UV region. We also consider the problem of hole self-trapping and calculate the hole self-trap energy.
    
BaZrO$_3$ is an ideal perovskite with space group Pm$\bar{3}$m and lattice constant of 4.19 \AA\cite{LEVIN2003170}, figure \ref{fig:struct}(a). Ba$_2$CaTeO$_6$ is a face-centered cubic double perovskite with space group Fm$\bar{3}$m and lattice constant of 8.3536 \AA, figure \ref{fig:struct}(b); its structure was determined by Fu \textit{et al.}\cite{FU20082523} and recently refined by Weil\cite{Weil:pj2054}. Ba$_2$K$_2$Te$_2$O$_9$ is a hexagonal triple perovskite with space group P6$_3$/mmc and lattice constants $a = 6.047$ \AA and $c = 16.479$ \AA, figure \ref{fig:struct}(c); it has been recently synthesized by Weil\cite{Weil:pj2054}. 
    
Electronic structure and optical properties calculations are carried out using the all-electron, full potential, linear augmented plane wave method as implemented in the WIEN2k code\cite{wien2k}. Each atom is surrounded by a muffin-tin sphere inside which the valence electrons wave function is expanded in spherical harmonics. In the interstitial region outside the muffin-tin spheres, the wave function is expanded in plane waves with a wave vector cutoff $K_{max}$ such that $R_{mt}K_{max} = 7$, where $R_{mt}$ is the smallest of the muffin-tin radii. Charge density is Fourier expanded up to $G_{max} = 12$ $Ry^{1/2}$. The calculations are based on the hybrid-functional HSE06\cite{doi:10.1063/1.1564060} potentials.

\begin{figure*}[ht]
\centering
\includegraphics[scale=0.25]{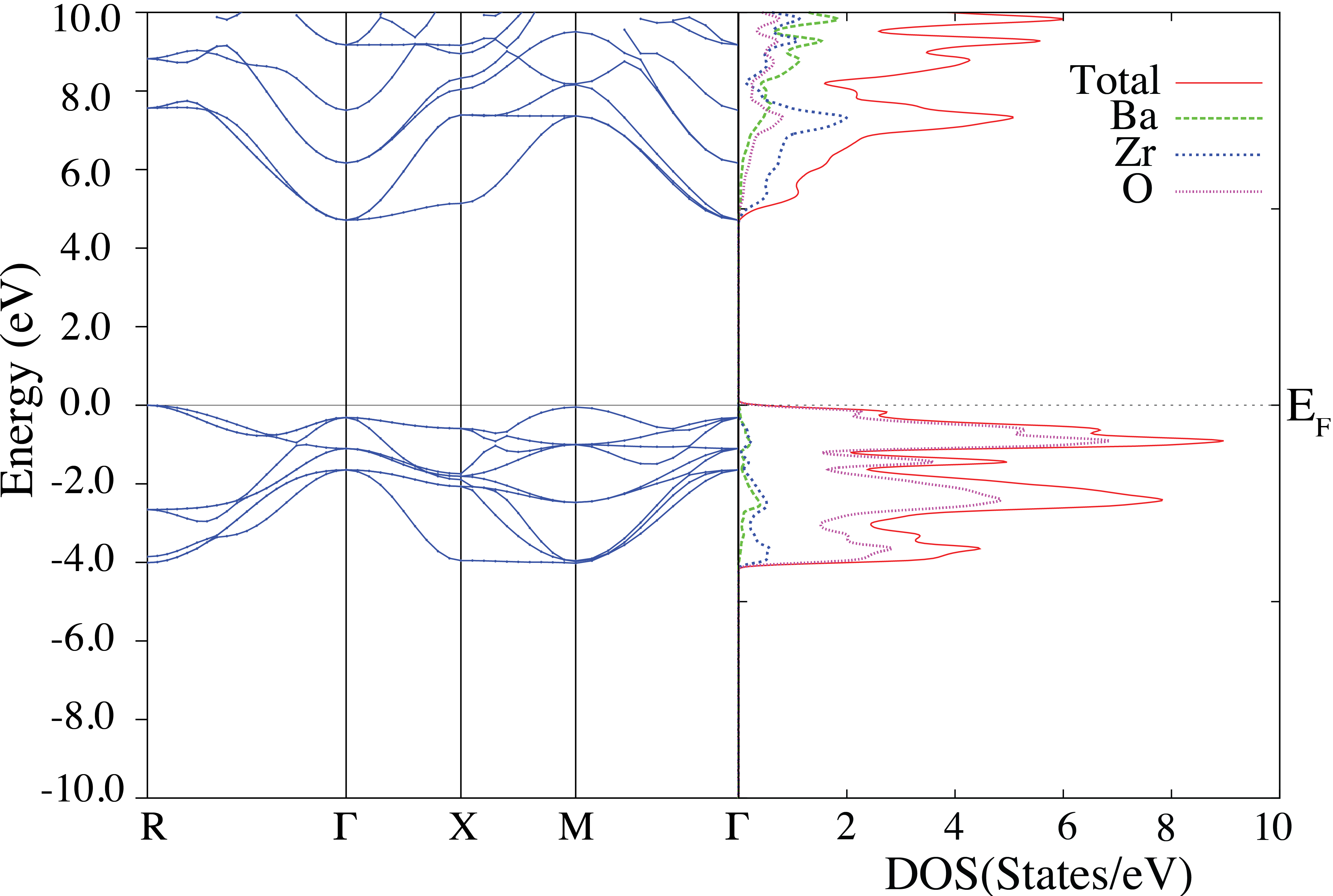}
\caption{Left panel: DFT/HSE computed electronic band dispersion of BaZrO$_3$ crystal along high symmetry directions in the first Brillouin zone. Right panel: Total and partial densities of states for Ba, Zr, and O atoms. The Fermi level has been set to zero matching the top of the valence band.}
\label{fig:BZO}
\end{figure*}

Our calculations for hole trapping use the HSE06 potential as implemented in the VASP code\cite{PhysRevB.54.11169}. Projector augmented wave potentials are used for the interaction between ionic cores and valence electrons. To study a small polaron, optimized supercells are used as starting structures. For the cubic BaZrO$_3$, a 3x3x3 supercell with 135 atoms is adopted, and one special k-point (0.25, 0.25, 0.25) is used to integrate over the Brillouin zone of the supercell. For the fcc double perovskite Ba$_2$CaTeO$_6$, a supercell of 80 atoms, obtained by doubling the conventional unit cell, is used, with a 4x4x2 k-mesh. As for the hexagonal triple perovskite Ba$_2$K$_2$Te$_2$O$_9$, a 2x2x1 supercell with 120 atoms is used, with a k-mesh of 2x2x2. Energy cutoff was set at 400 eV throughout these calculations.

\begin{figure*}[ht]
\centering
\includegraphics[scale=0.25]{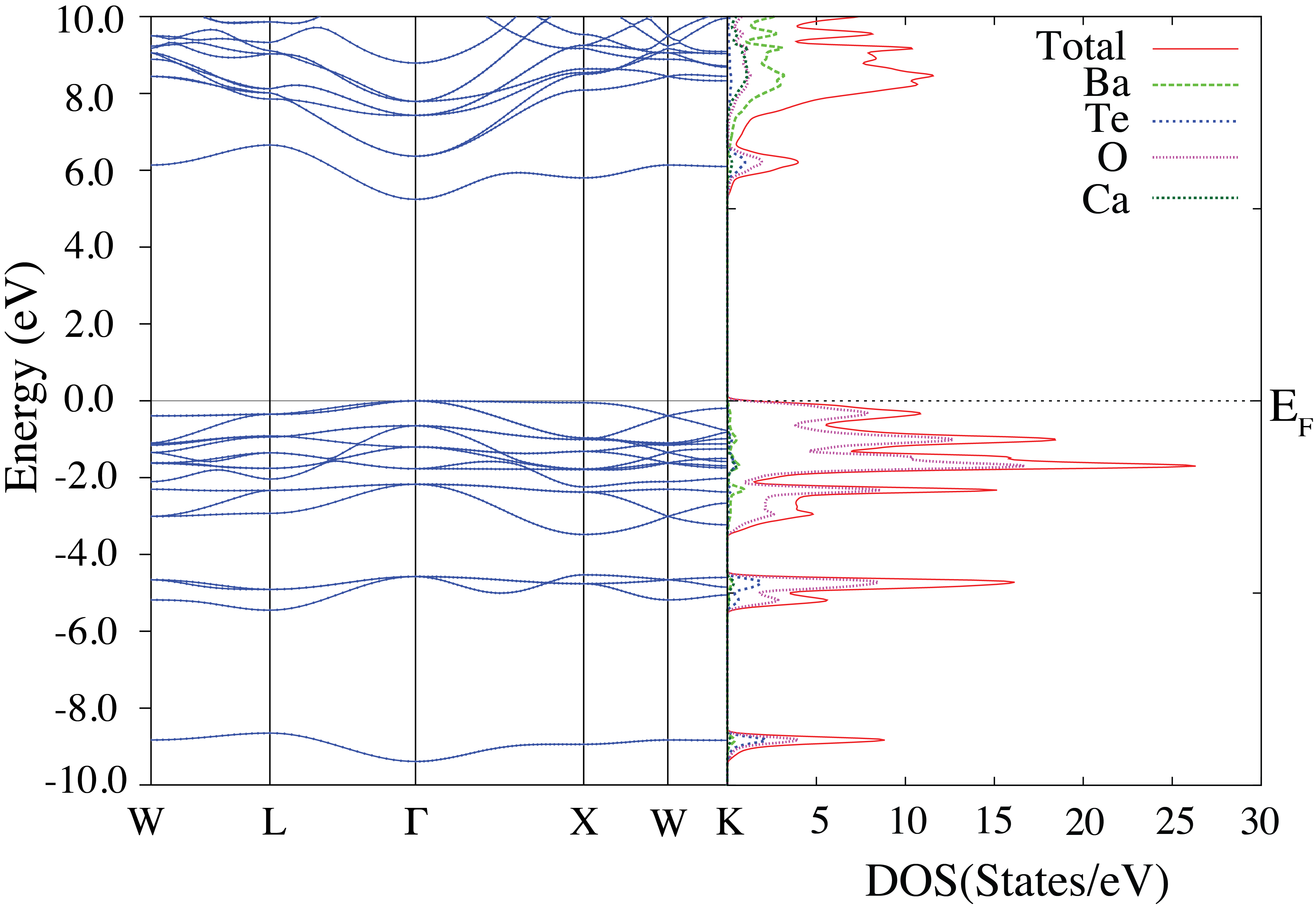}
\caption{Left panel: DFT/HSE computed electronic band dispersion of Ba$_2$CaTeO$_6$ crystal along high symmetry directions in the first Brillouin zone. Right panel: Total and partial densities of states for Ba, Te, O, and Ca atoms. The Fermi level has been set to zero matching the top of the valence band.}
\label{fig:bcto}
\end{figure*}

\begin{figure*}[ht]
\centering
\includegraphics[scale=0.25]{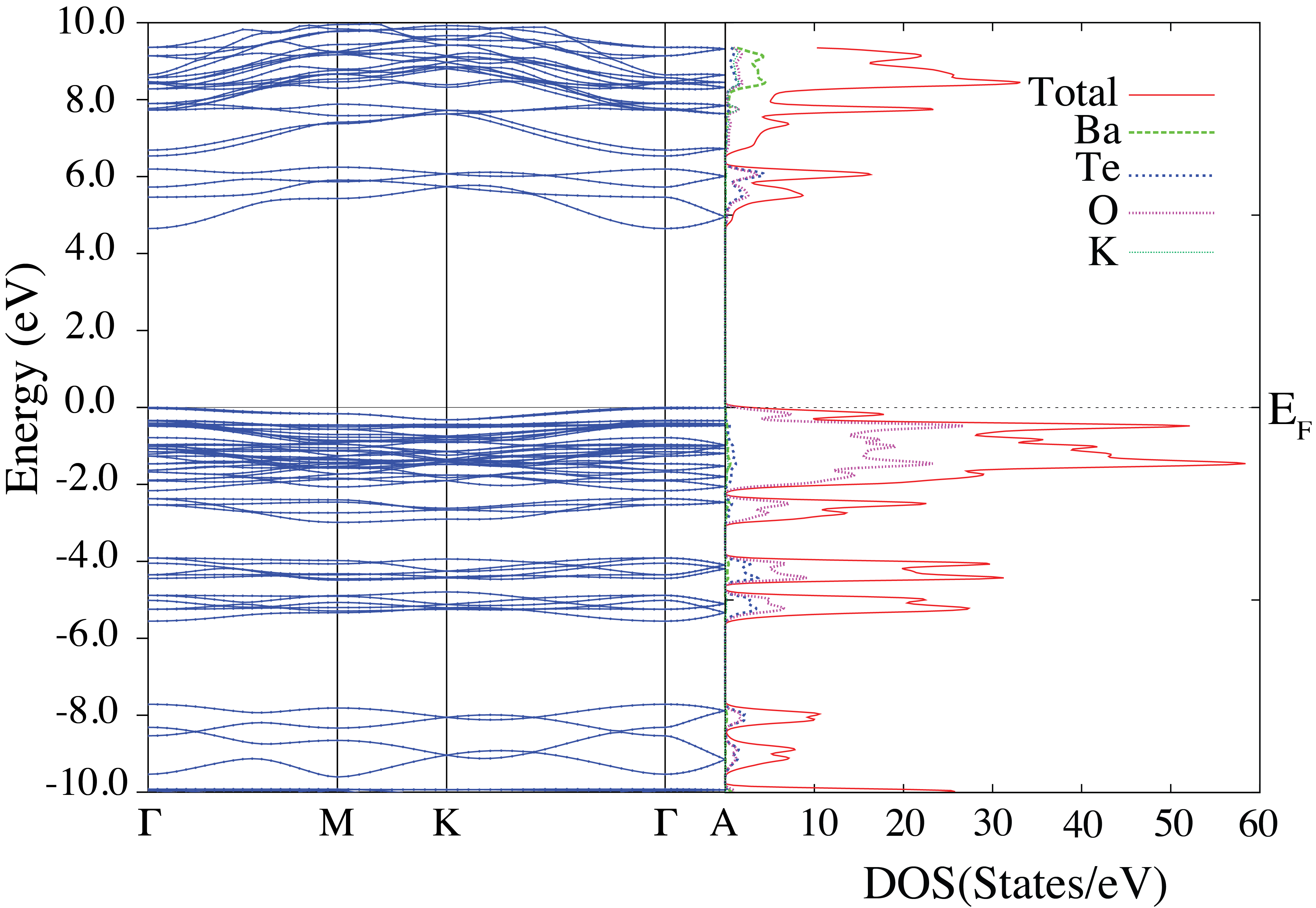}
\caption{Left panel: DFT/HSE computed electronic band dispersion of Ba$_2$K$_2$Te$_2$O$_9$ crystal along high symmetry directions in the first Brillouin zone. Right panel: Total and partial densities of states for Ba, Te, O, and K atoms. The Fermi level has been set to zero matching the top of the valence band.}
\label{fig:bkto}
\end{figure*}
    
The energy bands and densities of states of the single, double, and triple perovskites, calculated with the HSE06 potential, are shown in figures \ref{fig:BZO}, \ref{fig:bcto}, and \ref{fig:bkto}, respectively. The bandgaps, obtained with the PBE\cite{PhysRevLett.77.3865} and HSE06 potentials are presented in Table \ref{tab:gaps}. The calculated bandgap of BaZrO$_3$, 4.90 eV, is in good agreement with reported experimental value, 4.8-5.3 eV\cite{C4CY01201A,doi:10.1116/1.591472}. No experimental values for the bandgap are reported for the double and triple perovskites considered in this work.

 In the three crystals considered in this work, the upper valence bands are dominated by O-2p orbitals; these orbitals also contribute in a significant way to the lowest conduction band. Other contributions to the lowest conduction band are made by Zr-4d orbitals in the case of BaZrO$_3$ and by Te-5s orbitals in the case of the double and triple perovskites.

\begin{table*}[ht]
    \centering
    \caption{Calculated bandgaps using PBE and HSE06 potentials, and hole trapping energy E$_{ST}$}
    \label{tab:Table I}
    \begin{tabular}{c | c | c | c | c}
    \hline
    Crystal & PBE bandgap (eV) &  HSE06 bandgap (eV) &  Bandgap type & E$_{ST}$ (eV) \\
    \hline
    BaZrO$_3$ & 2.94    & 4.90 & Indirect & 0.247   \\
    Ba$_2$CaTeO$_6$ & 3.43     & 5.24 & Direct & 0.256 \\ 
    Ba$_2$K$_2$Te$_2$O$_9$ & 2.88 & 4.65 & Direct & 0.248 \\
    \hline
    \end{tabular}
    \label{tab:gaps}
    \end{table*}

\begin{figure*}[ht]
\centering
\includegraphics[scale=0.25]{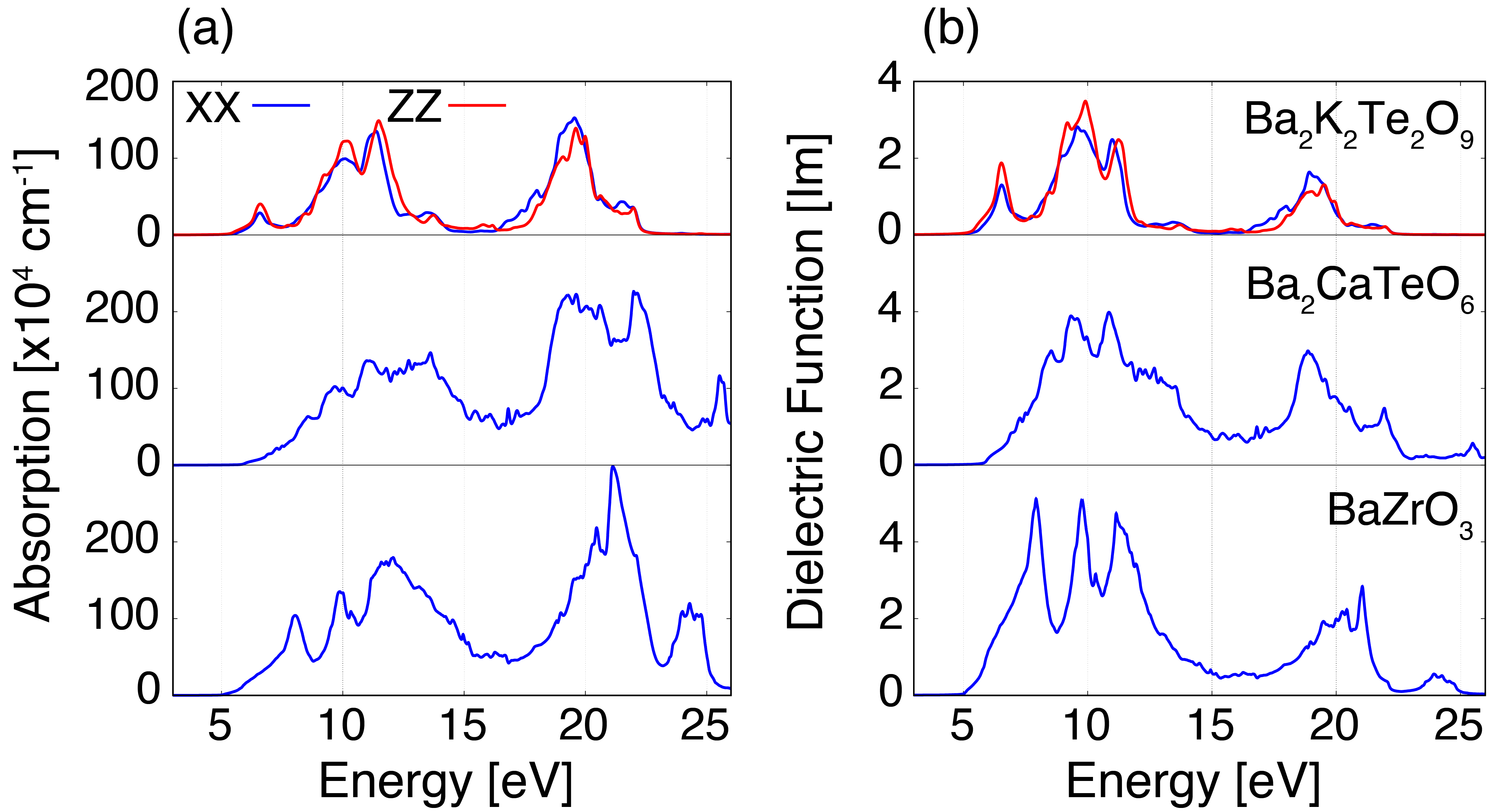}
\caption{(a) Optical absorption and (b) Imaginary part of the dielectric function computed from hybrid functionals for Ba$_2$K$_2$Te$_2$O$_9$ (top), Ba$_2$CaTeO$_6$ (middle) and BaZrO$_3$ (bottom). Note that Ba$_2$K$_2$Te$_2$O$_9$ has a hexagonal structure, hence the X and Z components of the absorption coefficient and dielectric function are not equal.}
\label{fig:abs}
\end{figure*}

 In figure \ref{fig:abs} we present the calculated absorption coefficients and the imaginary parts of the dielectric functions for the three crystals. Note that since the triple perovskite is hexagonal, $\epsilon_x = \epsilon_y \neq \epsilon_z$. Similarly, the absorption coefficient for waves polarized in the a-b plane differs from that for waves polarized in the c-direction. The peaks in the absorption spectra and the imaginary part of the dielectric function result from transitions from O-\textit{2p} orbitals in the valence band to Zr-\textit{4d} orbitals in BaZrO$_3$ or to Te-\textit{5s} orbitals in BaCaTeO$_6$ and Ba$_2$K$_2$Te$_2$O$_9$.
 
 To calculate the hole self-trapping energy, an electron is removed from the supercell and the total energy $E_1$ is calculated using hybrid-functional density functional theory. Removal of an electron from the perfect supercell results in a delocalized hole at the valence band maximum. Next, atoms surrounding a certain oxygen atom are given random displacements, the atomic positions are relaxed, and the total energy $E_2$ is calculated. An energy $E_2$ lower than $E_1$ indicates that the hole is self-trapped; $E_1 - E_2$ would then correspond to the hole self-trapping energy. In all three crystals, we find that the hole is stabilized by small lattice distortion around an oxygen atom. In Table I we report the hole self-trapping energy for each of the perovskites considered in this work. Self-trapping of holes points to the difficulty of achieving p-type doping in these materials unless techniques similar to those employed by Pandey \textit{et al.}\cite{PhysRevMaterials.3.053401} are implemented.
 
 To summarize, in this letter we studied the electronic and optic features of three barium-based perovskites from first principles calculations based on hybrid functionals. Two of them, Ba$_2$CaTeO$_6$ and Ba$_2$K$_2$Te$_2$O$_9$, are proposed as novel ultrawide bandgap semiconductors with bandgaps of 5.24 eV and 4.65 eV, respectively. Contrary to ideal perovskites where the bandgap exhibit an indirect transition from $R$ in the valance band to $\Gamma$ in the conduction band\cite{C5TC04172D}, we find that the double and triple perovskites show a direct bandgap transition at the $\Gamma$-point. Similarly to other well established UWBG semiconductors, we predict that hole self-trapping is favorable through lattice distortions near oxygen sites. We hope that this work is viewed as a proof of concept and that it would encourage further experimental and theoretical investigations into perovskites as UWBG semiconductors, especially those exhibiting double and triple perovskite structures which have not been widely studied.

\begin{acknowledgements}
This work was supported in part by the National Science Foundation under PREM grant no. DMR-1523588 and CREST grant no. HRD-1547723.
\end{acknowledgements}

\textit{The data that support the findings of this study are available from the corresponding author upon reasonable request}

\bibliography{references.bib}

\end{document}